\documentclass[preprint2]{aastex}

\def  \be	 {\begin{equation}}
\def  \ee	 {\end{equation}}
\def  \beq	 {\begin{eqnarray}}
\def  \eeq	 {\end{eqnarray}}
\def  \bfK	 {\mbox{\boldmath${\cal K}$}}
\def  \bfG	 {\mbox{\boldmath${\cal G}$}}

\begin{document}

\title{\bf Evolution of magnetic field curvature in the Kulsrud-Anderson dynamo theory}
\author{Leonid Malyshkin}
\affil{Princeton University Observatory, Princeton NJ 08544, USA}
\email{leonmal@astro.princeton.edu}
\date{\today}

\begin{abstract}
We find that in the kinematic limit the ensemble averaged square of the curvature of 
magnetic field lines is exponentially amplified in time by the turbulent motions in a highly 
conductive plasma. At the same time, the ensemble averaged curvature vector exponentially 
decays to zero. Thus, independently of the initial conditions, the fluctuation field 
becomes very curved, and the curvature vector becomes highly isotropic.
\end{abstract}

\keywords{ISM: magnetic fields --- MHD --- turbulence --- methods: analytical}


It was shown by Kulsrud and Anderson (1992) that MHD turbulent dynamo action builds magnetic field energy 
primarily on scales smaller than the smallest turbulent eddy size (which is the viscosity scale), but
still larger than the resistive scale (provided the magnetic Prandtl number is large). This result was 
found under assumption that the ``kinematic'' approximation is valid, i.e.~the field is weak enough that 
it does not affect the turbulent motions. In this paper we calculate the evolution of magnetic field
curvature within the framework of the Kulsrud-Anderson kinematic dynamo theory. Such calculations are of 
great interest because a rapid built up of curvature may quickly break down the kinematic approximation 
and change the dynamo action on very small scales considerably.
 
Following Kulsrud and Anderson (1992) we make the following assumptions. We use the ``kinematic'' 
approximation. We neglect resistivity (infinite magnetic Prandtl number limit). We assume that the 
turbulence is incompressible, homogeneous, isotropic and static, and we use zero correlation time 
approximation for the turbulent motions:
\beq
&&V_\alpha(t,{\bf r})=\frac{1}{(2\pi)^3}\int V_\alpha(t,{\bf k})\,e^{i{\bf k}\cdot{\bf r}}\,d^3{\bf k},
\label{V}
\\
&&\langle V_\alpha(t,{\bf k})\rangle=0,
\qquad
\label{Vk}
\\
&&\langle V^*_\alpha(t',{\bf k'})V_\beta(t,{\bf k})\rangle 
=[J_k(\delta_{\alpha\beta}-{\hat k}_\alpha{\hat k}_\beta)
\qquad
\nonumber\\
&&\qquad {}+i{\bar J}_k\varepsilon_{\alpha\gamma\beta}k_\gamma]\delta({\bf k'}-{\bf k})\delta(t'-t).
\qquad
\label{Vk_Vk}
\eeq
Here and below $\langle ... \rangle$ means ensemble average, $\delta_{\alpha\beta}$ is the Kronecker symbol,
$\varepsilon_{\alpha\beta\gamma}$ is the unit anti-symmetric tensor, $\delta(t'-t)$ and
$\delta({\bf k'}-{\bf k})$ are the Dirac $\delta$-functions, ${\bf {\hat k}}={\bf k}/k$ is a unit vector, 
and we always assume summation over repeated indices. Functions $J_k$ and ${\bar J}_k$ are the normal 
and the helical parts of the turbulence, they depend only on the absolute value of $\bf k$. 
We make no assumptions about the statistics of initial magnetic field.

Using equations~(\ref{V}) and~(\ref{Vk_Vk}), it is straightforward to calculate the correlation 
tensors between velocities and their spatial derivatives, taken at the same point of space 
$\bf r$ but at different time, $t$ and $t'$:
\beq
&&\langle V_{\alpha}(t',{\bf r}) V_{\beta}(t,{\bf r})\rangle 
= (\eta_T/2\pi)\delta_{\alpha\beta}\delta(t'-t),
\nonumber
\\
&&\left\langle V_{\alpha} V_{\beta ; \gamma}\right\rangle
= \alpha\varepsilon_{\alpha\beta\gamma}\delta(t'-t),
\nonumber
\\
&&\left\langle V_{\alpha ; \beta} V_{\gamma ; \delta}\right\rangle
= (\gamma/5)(5\delta_{\alpha\gamma}\delta_{\beta\delta}-\delta_{\alpha\beta\gamma\delta})\delta(t'-t),
\nonumber
\\
&&\left\langle V_{\alpha} V_{\beta ; \gamma\delta}\right\rangle
= -\left\langle V_{\alpha ; \delta} V_{\beta ; \gamma}\right\rangle,
\nonumber
\\
&&\left\langle V_{\alpha ; \beta} V_{\gamma ; \delta\tau}\right\rangle
= \omega\varepsilon_{\alpha\gamma\eta}\delta_{\eta\beta\delta\tau}\delta(t'-t),
\label{V_V}
\\
&&\left\langle V_{\alpha} V_{\beta ; \gamma\delta\tau}\right\rangle
= -\left\langle V_{\alpha ; \tau} V_{\beta ; \gamma\delta}\right\rangle,
\nonumber
\\
&&\left\langle V_{\alpha ; \beta\gamma} V_{\delta ; \tau\eta}\right\rangle
= (\lambda/15)(7\delta_{\alpha\delta}\delta_{\beta\gamma\tau\eta}
\nonumber
\\ 
&&\qquad\qquad\qquad {}-\delta_{\alpha\beta\gamma\delta\tau\eta})\delta(t'-t).
\nonumber
\eeq
Here and below, in order to shorten notations, spatial derivatives are assumed to be taken with 
respect to all indices that are listed after ``$\,;\,$'' signs, 
e.g.~$V_{\alpha ; \beta}=\partial V_{\alpha}/\partial x_\beta$, 
$V_{\alpha ; \beta\gamma}=\partial^2V_{\alpha}/\partial x_\beta\partial x_\gamma$.
We have also introduced the symmetric tensors
\beq
\begin{array}{l} 
\delta_{\alpha\beta\gamma\delta}
=\delta_{\alpha\beta}\delta_{\gamma\delta}+\delta_{\alpha\gamma}\delta_{\beta\delta}
+\delta_{\alpha\delta}\delta_{\beta\gamma},
\\
\delta_{\alpha\beta\gamma\delta\tau\eta}
=\delta_{\alpha\beta}\delta_{\gamma\delta\tau\eta}+\delta_{\alpha\gamma}\delta_{\beta\delta\tau\eta}
+\delta_{\alpha\delta}\delta_{\beta\gamma\tau\eta}
\\ 
\qquad\qquad\qquad {}+\delta_{\alpha\tau}\delta_{\beta\gamma\delta\eta}
+\delta_{\alpha\eta}\delta_{\beta\gamma\delta\tau},
\end{array}
\label{TENSORS}
\eeq
and the constants (Kulsrud \& Anderson 1992)
\beq
\eta_T &=& \frac{4\pi}{3}\frac{1}{(2\pi)^6}\int J_k \:4\pi k^2dk,
\nonumber
\\
\alpha &=& \frac{1}{3}\frac{1}{(2\pi)^6}\int k^2 {\bar J}_k \:4\pi k^2dk,
\nonumber
\\
\gamma &=& \frac{1}{3}\frac{1}{(2\pi)^6}\int k^2 J_k \:4\pi k^2dk,
\label{CONSTANTS}
\\
\omega &=& \frac{1}{15}\frac{1}{(2\pi)^6}\int k^4 {\bar J}_k \:4\pi k^2dk,
\nonumber
\\
\lambda &=& \frac{1}{7}\frac{1}{(2\pi)^6}\int k^4 J_k \:4\pi k^2dk.
\nonumber
\eeq

The equation for the evolution of magnetic field $\bf B$ in an incompressible highly conductive
fluid (Landau \& Lifshitz 1983) is
\beq
\partial B_\alpha/\partial t = V_{\alpha ; \beta} B_\beta-V_\beta B_{\alpha ; \beta}.
\nonumber
\eeq
We use this formula to derive the equation for the evolution of the unit vector ${\bf b}={\bf B}/B$ 
tangential to magnetic field lines:
\beq
\partial b_\alpha/\partial t = V_{\alpha ; \beta} b_\beta
-V_{\beta ; \gamma} b_\alpha b_\beta b_\gamma-V_\beta b_{\alpha ; \beta}.
\label{b_EQ}
\eeq
Now, for a given point of space we assume $\bf b$ is known at zero time, $t=0$, and we solve 
this equation by iterating it twice in time, similar to the calculations of Kulsrud and 
Anderson (1992). Considering $t>0$ as an expansion parameter, we 
have
\beq
{\bf b}(t)={}^0{\bf b}+{}^1{\bf b}(t)+{}^2{\bf b}(t),
\label{EXPANSION}
\eeq
where ${}^0{\bf b}$ is the value of $\bf b$ at zero time, ${}^0{\bf b}={\bf b}(0)$, and 
${}^1{\bf b}(t)\propto t$ and ${}^2{\bf b}(t)\propto t^2$ are the first and the second order 
corrections to ${\bf b}(t)$, obtained by iterating equation~(\ref{b_EQ}) twice in time,
\beq
&& {}^{1\!}b_\alpha(t) = \int_0^t \! \left\{V_{\alpha ; \beta}(t''){}^{0\!}b_\beta
-V_{\beta ; \gamma}(t'') {}^{0\!}b_\alpha {}^{0\!}b_\beta {}^{0\!}b_\gamma\right.
\nonumber
\\
&& \qquad\qquad\quad \left. {}-V_\beta(t'') {}^{0\!}b_{\alpha ; \beta}\right\} dt'',
\label{b_1}
\\
&& {}^{2\!}b_\alpha(t) = \int_0^t \! \left\{V_{\alpha ; \beta}(t'){}^{1\!}b_\beta(t')
-V_{\beta ; \gamma}(t')\times{} \right. 
\nonumber
\\
&& \: {}\times\left[{}^{1\!}b_\alpha(t') {}^{0\!}b_\beta {}^{0\!}b_\gamma
+{}^{0\!}b_\alpha {}^{1\!}b_\beta(t') {}^{0\!}b_\gamma\right.
\nonumber
\\
&& \left.\left. {}+{}^{0\!}b_\alpha {}^{0\!}b_\beta {}^{1\!}b_\gamma(t')\right]
-V_\beta(t') {}^{1\!}b_{\alpha ; \beta}(t')\right\} dt'.
\label{b_2}
\eeq

The {\it ensemble averaged} curvature 
$\mbox{\boldmath${\cal K}$}\equiv \big\langle({\bf b}\nabla){\bf b}\big\rangle$, 
up to second order terms, is
\beq
&&{\cal K}_\alpha(t) = \big\langle b_\beta(t)b_{\alpha ; \beta}(t)\big\rangle
= \big\langle{}^{0\!}b_\beta {}^{0\!}b_{\alpha ; \beta}\big\rangle
\nonumber
\\
&&\; {}+ \big[ 
\big\langle{}^{2\!}b_\beta\big\rangle {}^{0\!}b_{\alpha ; \beta} 
+ {}^{0\!}b_\beta {\big\langle{}^{2\!}b_\alpha\big\rangle}_{ ; \beta}
+ \big\langle{}^{1\!}b_\beta {}^{1\!}b_{\alpha ; \beta}\big\rangle
\big]
\nonumber
\\
&&\; {}= {}^{0\!}{\cal K}_\alpha + t\,\big\{\!-(7\gamma/5){}^{0\!}{\cal K}_\alpha
- (4\gamma/5)\big\langle{}^{0\!}b_\alpha {}^{0\!}b_{\beta ; \beta}\big\rangle
\nonumber
\\
&&\; {}+ \alpha \varepsilon_{\alpha\beta\gamma}{}^{0\!}{\cal K}_{\gamma ; \beta} 
+ (\eta_T/4\pi){}^{0\!}{\cal K}_{\alpha ; \beta\beta} \big\}.
\label{K}
\eeq
Here, we use expansion~(\ref{EXPANSION}) to find $\bfK(t)$. The zero order term is 
${}^{0\!}{\cal K}_\alpha=\big\langle{}^{0\!}b_\beta {}^{0\!}b_{\alpha ; \beta}\big\rangle$,
all first order terms average out according to formula~(\ref{Vk}), and the second 
order terms are given in brackets $[...]$ on the second line of 
equation~(\ref{K}).\footnote{
Note that we use $\langle{}^{2\!}b_{\alpha ; \beta}\rangle={\langle{}^{2\!}b_\alpha\rangle}_{ ; \beta}$, 
which follows from important identity $\langle (V_{\alpha ; \beta_1...\beta_n} 
V_{\gamma ; \delta_1...\delta_m})_{; \tau_1...\tau_p}\rangle\equiv 0$. This reflects 
$\delta({\bf k'}-{\bf k})$ correlation property of the turbulence, see eq.~(\ref{Vk_Vk}). Thus, ensemble 
averaging and taking derivatives can be exchanged for any second order quantity.\label{FOOTNOTE_1}
}
The final result for ${\cal K}_\alpha(t)$ was obtained by making use of 
equations~(\ref{V_V}),~(\ref{b_1}) and~(\ref{b_2}). 

Introducing another ensemble averaged vector 
$\bfG\equiv \big\langle{\bf b}\,{\rm div}\,{\bf b}\big\rangle$, we write 
the differential equations for time evolution of $\bfK(t)$ and $\bfG(t)$ as
\beq
\frac{\partial\bfK}{\partial t} = - \frac{7\gamma}{5}\bfK
- \frac{4\gamma}{5}\bfG + \alpha(\nabla\times\bfK) + \frac{\eta_T}{4\pi}\triangle\bfK,
\label{K_EQ}
\\
\frac{\partial\bfG}{\partial t} = - \gamma\bfK - \frac{8\gamma}{5}\bfG 
+ \alpha(\nabla\times\bfG) + \frac{\eta_T}{4\pi}\triangle\bfG.
\;\;\;\:
\label{G_EQ}
\eeq
Here, the first equation directly follows from formula~(\ref{K}), and the second equation 
for the evolution of $\bfG$ can be found by calculations similar to the calculations that led to 
equation~(\ref{K}), i.e.~by expanding $\bfG(t)$ up to second order terms, by making use of 
equations~(\ref{b_1}),~(\ref{b_2}) and of equations~(\ref{V_V}) to carry out ensemble 
averaging.

In a similar way, and after considerable algebra, we find the differential equations describing 
the evolution of the ensemble averaged square of the curvature, 
${\cal K}^2\equiv \big\langle{[({\bf b}\nabla){\bf b}]}^2\big\rangle \ne \bfK^2$,
\beq
\frac{\partial {\cal K}^2}{\partial t} = \frac{16\gamma}{5}{\cal K}^2 + \frac{8\gamma}{5}D_{\cal K}
+ \frac{\eta_T}{4\pi}\triangle {\cal K}^2 + \frac{12\lambda}{5},
\label{K2_EQ}
\\
\frac{\partial D_{\cal K}}{\partial t} = -\frac{7\gamma}{5}D_{\cal K} - \frac{4\gamma}{5}D_{\cal G}
+ \frac{\eta_T}{4\pi}\triangle D_{\cal K},
\qquad\;\,
\label{DK_EQ}
\\
\frac{\partial D_{\cal G}}{\partial t} = -\gamma D_{\cal K} - \frac{8\gamma}{5}D_{\cal G}
+ \frac{\eta_T}{4\pi}\triangle D_{\cal G}.
\qquad\quad\;\,
\label{DG_EQ}
\eeq
Here, the differential equations for ensemble averaged scalars 
$D_{\cal K}\equiv \big\langle {\rm div}\,[({\bf b}\nabla){\bf b}]\big\rangle = {\rm div}\,\bfK$ and 
$D_{\cal G}\equiv \big\langle {\rm div}\,[{\bf b}\,{\rm div}\,{\bf b}]\big\rangle = {\rm div}\,\bfG$
can be found by taking the divergence of equations~(\ref{K_EQ}) and~(\ref{G_EQ}) (see the first 
footnote on page~\pageref{FOOTNOTE_1}).

We can drop $\alpha(\nabla\times\bfK)$ and $\alpha(\nabla\times\bfG)$ terms in equations~(\ref{K_EQ}) 
and~(\ref{G_EQ}) because the helical part of the turbulence, ${\bar J}_k$, is usually negligible on
scales of the smallest turbulent eddy.\footnote{
E.g.~in a galaxy, if the turbulence is Kolmogoroff, ${\bar J}_k\propto k^{-7}$ and $J_k\propto k^{-13/3}$
(Kulsrud \& Anderson 1992), these smallest turbulent scales make the largest contributions to $\alpha$ 
and $\gamma$, see eqs.~(\ref{CONSTANTS}).
} 
In this case equations~(\ref{K_EQ})--(\ref{DG_EQ}) have the following solutions:
\beq
\!\left[\!\!\begin{array}{c}\bfK\\ \bfG\end{array}\!\!\right]\!
={\bf Q}_1 \!\left[\!\!\!\begin{array}{c}1\\-1\end{array}\!\!\!\right]\! e^{-3\gamma t/5} 
+{\bf Q}_2 \!\left[\!\!\begin{array}{c}4\\5\end{array}\!\!\right]\! e^{-12\gamma t/5},
\nonumber 
\\
\!\left[\!\!\begin{array}{c}{\cal K}^2\\D_{\cal K}\\D_{\cal G}\end{array}\!\!\right]\!
=Q_3 \!\left[\!\!\begin{array}{c}1\\0\\0\end{array}\!\!\right]\! e^{16\gamma t/5} 
+Q_4 \!\left[\!\!\!\begin{array}{c}8\\-19\\19\end{array}\!\!\!\right]\! e^{-3\gamma t/5}
\nonumber
\\
{}+Q_5 \!\left[\!\!\!\begin{array}{c}-8\\28\\35\end{array}\!\!\!\right]\! e^{-12\gamma t/5}
+\frac{3\lambda}{4\gamma} \!\left[\!\!\begin{array}{c}1\\0\\0\end{array}\!\!\right]\! 
\big(e^{16\gamma t/5}-1\big).
\nonumber 
\eeq
Here, the functions $Q_i(t,{\bf r})$, $i=1,2,3,4,5$, are the solutions of the same simple 
diffusion equation
\beq
\frac{\partial Q_i}{\partial t} = \frac{\eta_T}{4\pi}\triangle Q_i(t,{\bf r}), \quad i=1,2,3,4,5
\label{DIFFUSION_EQ}
\eeq
with different initial conditions:
\beq
&&{\bf Q}_1(0,{\bf r})=(1/9){\big[5\bfK-4\bfG\big]}_{t=0},
\nonumber
\\
&&{\bf Q}_2(0,{\bf r})=(1/9){\big[\bfK+\bfG\big]}_{t=0}.
\nonumber
\\
&&Q_3(0,{\bf r})={\big[{\cal K}^2+(1/133)(48D_{\cal K}-8D_{\cal G})\big]}_{t=0},
\nonumber
\\
&&Q_4(0,{\bf r})=(1/171){\big[-5D_{\cal K}+4D_{\cal G}\big]}_{t=0},
\nonumber
\\
&&Q_5(0,{\bf r})=(1/63){\big[D_{\cal K}+D_{\cal G}\big]}_{t=0},
\nonumber
\eeq
We see that the ensemble averaged square of the curvature, ${\cal K}^2$, exponentially grows with rate 
$16\gamma/3$. This is even faster than the rate of magnetic field energy growth, $2\gamma$ (Kulsrud \& 
Anderson 1992). At the same time, the ensemble averaged curvature vector, $\bfK$, exponentially decays 
with rate $-3\gamma/5$. Therefore, $\bfK$ becomes highly isotropic.
According to diffusion equation~(\ref{DIFFUSION_EQ}), we find that ${\cal K}^2$ and $\bfK$ become 
homogeneous on scales $L$ after a diffusion time $\sim 4\pi L^2/\eta_T$.

Let consider the case when the initial magnetic field is constant in space, $\bf B={\rm const}$ and 
$\bf b={\rm const}$. Then $Q_i(t,{\bf r})=0$, $i=1,2,3,4,5$, and we find that  
${\cal K}^2=(3\lambda/4\gamma)\big(e^{16\gamma t/5}-1\big)$. 
Therefore, even if there is no initial curvature, the field quickly becomes very curved. The curvature 
first develops linearly in time, ${\cal K}^2\approx 12\lambda t/5$, because of second order 
spatial derivatives of the turbulent velocities, represented by the ``battery'' term $12\lambda/5$ in 
equation~(\ref{K2_EQ}). Second, at time $t\sim 5/16\gamma$ the averaged curvature reaches the smallest
turbulent eddy scale size, $l_{\rm eddy}\sim \gamma/\lambda$, and afterwards exponentiates rapidly.

It was recently suggested by Steven Cowley, on the basis of numerical simulations, that on 
scales smaller than the smallest turbulent eddy the turbulent motions can only stretch magnetic field 
lines. It was argued that the rapid exponential increase of the fluctuation field wavenumber 
$k=(k_\perp^2+k_\parallel^2)^{1/2}$ found by Kulsrud and Anderson (1992) is mainly due to the increase 
of $k_\perp$, the wave number perpendicular to the magnetic field lines, while the parallel wave number 
$k_\parallel$ stays approximately equal to the smallest eddy size, so that $k_\perp\gg k_\parallel$. 
The results we obtained, that the ensemble (volume) averaged curvature squared, ${\cal K}^2$, increases
exponentially in time, could be consistent with the folding nature of small-scale fields. For example,
the curvature could be comparable to the smallest turbulent eddy size scale over the bulk of the volume, 
and be enormously large over the remaining small region of the volume (Schekochihin \& Cowley 2001). 

The consequence of the rapid development of curvature on size scales smaller than the smallest turbulent 
eddy size scale is that the Lorentz force and viscosity stress on these very small scales become important 
before energy equipartition, between the field and the turbulence, occurs on the eddy scale. Thus, the 
magnetic energy at these very small scales may be suppressed at least to some extent.

The author is very grateful to Russell Kulsrud for suggesting this problem and for many interesting 
and extremely fruitful discussions of it. The author would like to thank Alexander Schekochihin and 
Steven Cowley for extensive discussions of the problem and for insightful comments. The author is 
especially grateful to Bruce Draine for financial support under NSF grant AST-9988126.



\end{document}